\input{aipcheck}

\documentclass[
    ,final            
  ]
  {aipproc}

\usepackage{aas_macros}
\layoutstyle{6x9}

\begin{document}

\title{Flickering in Black Hole Accretion discs}

\classification{95.30.Qd,97.10.Gz,97.60.Lf,97.80.Jp,98.54.-h}
\keywords      {Magnetohydrodynamics and plasmas,Accretion and accretion disks,Black holes,X-ray binaries,Quasars; active or peculiar galaxies, objects, and systems}

\author{M. Mayer}{
  address={Institute of Astronomy, Madingley Road, Cambridge CB3 0HA, UK}
}

\author{J.E. Pringle}{
  address={Institute of Astronomy, Madingley Road, Cambridge CB3 0HA, UK}
}

\begin{abstract}
We present an extension of the \citet{2004MNRAS.348..111K} model for
the flickering of black hole accretion discs by taking proper account
for the thermal properties of the disc. 

First we develop a one-dimensional, vertically averaged, one-zone model for
an optically thick accretion disc and study the temporal
evolution. This limits the current model to the so-called high/soft
state, where the X-Ray spectrum is dominated by a thermal black-body component. 

Then we couple this disc model to the flickering process as described in
\citet{2004MNRAS.348..111K}. Thus we consider the evolution of a
poloidal magnetic field subject to a magnetic dynamo. 
By comparing to observations of X-Ray
binaries in the high-soft state, we can constrain the strength of the
energy density of the poloidal magnetic field to a few percent of the
energy density of the intrinsic disc magnetic field.
\end{abstract}

\maketitle

\section{Introduction}

Accretion powered X-Ray binaries, on both the galactic (AGN - Active
galactic nuclei) and stellar scale, display significant aperiodic
variability \cite[for the latest review, see][]{2006astro.ph..6352R}. 

Power density spectra (PDS) of X-Ray binaries show that
most variability is created on timescales of about seconds and
longer. This is evident by a $\nu^{-1}$ behaviour of the PDS at low
frequencies and a break to a steeper slope at frequencies above a few
Hz. This is incompatible with the predominant timescales close to
the black hole where most of the luminosity, i.e. variability, is released. Moreover, in
the standard disc picture, variability created further out is damped
on the viscous timescale long before it reaches the inner part of the
disc where it would be observable. 

Thus the origin of this variability is not well understood. There are
however more statistical relations characterising this
variability. 
\citet{2001MNRAS.323L..26U} finds a linear relationship
between the rms variability and the flux of the source. 
\citet{2002MNRAS.336..817M} added higher-order statistics by
introducing a characteristic shape of the so-called bicoherence.

\citet{2005MNRAS.359..345U} give a comprehensive summary of these and other statistical properties and a test of
several theoretical models against the observed quantities of
the flickering variability. They rule out many models and are left
with a class of models proposed by
\citet{1997MNRAS.292..679L}. He suggests that uncorrelated fluctuations in
the viscosity parameter $\alpha$ (as defined in \citet{1973A&A....24..337S}) on about
the viscous timescale can reproduce the observed shape of the power density spectra. 

\citet{2004MNRAS.348..111K} use this
phenomenological model to fill in a physical picture. They consider
the temporal evolution of a poloidal magnetic field that changes on about the dynamical timescale according to magnetic dynamo
theory and is subject to induction and radial advection. Although the
timescale for the change of the magnetic field locally is connected to
the dynamical timescale, they are able to reproduce the observed
timescales for flickering in X-Ray binaries. They use the result of
\citet{2003ApJ...593..184L} who find that the poloidal magnetic field
in neighbouring cells is sufficiently aligned on a timescale of
$2^{R/H}$ times the dynamical timescale, where $R$ is the radial
distance from the black hole and $H$ the local scale height of the
disc. This timescale is fairly long for geometric thin ($H/R\ll 1$)
discs. If there happens to be such an alignment, then magnetic torques
will enhance the accretion flow and eventually launch an large-scale
inflow. This then might eventually create a disk wind/outflow. 

\section{Our model} 

While the \citet{2004MNRAS.348..111K} model already reproduces the
statistical properties of the flickering, it contains a number of
simplifications: First, they use $H/R=$const. Thus they do not
account for the vertical structure of the accretion disc. While
$H/R=$const. is a good approximation in the gas pressure dominated
regime, in the radiation pressure dominated regime $H$ becomes
constant. Moreover, $H/R$ certainly will be a function of time and
radius, if there are limit cycles present. 

Secondly, since their discs are very thin ($H/R=0.08$), they
use a fairly high value for the maximum allowed energy density of the
poloidal magnetic field. They typically use a ten times larger value
than the local intrinsic magnetic field energy density.  

We started to extend this model to account for
a more realistic vertical structure and to trace the thermal and
viscous evolution of these discs in the presence of the flickering
mechanism. We introduce the basic physical ingredients of the model
below here but refer the reader to \citet{2006MNRAS.368..379M} for more
information on the details of the numerical implementation.

As mentioned before, the flickering in the \citet{2004MNRAS.348..111K}
model is produced locally by a magnetic dynamo process which through
reconnection can drive an enhanced mass inflow at times. The dynamoes
operate in grid cells of a characteristic width $\Delta R\approx
H$. Hence in cases where $H/R$ is no longer constant, we need to have
a grid which adapts to the changing conditions. We therefore
implemented a one-dimensional self-adaptive grid (a type of 1D adaptive
mesh refinement). The AMR is written that if a refinement/coarsening
occurs, both the mass, energy, angular momentum and magnetic flux is
conserved. 

\begin{figure}
  \centering
  \includegraphics[width=0.7\textwidth]{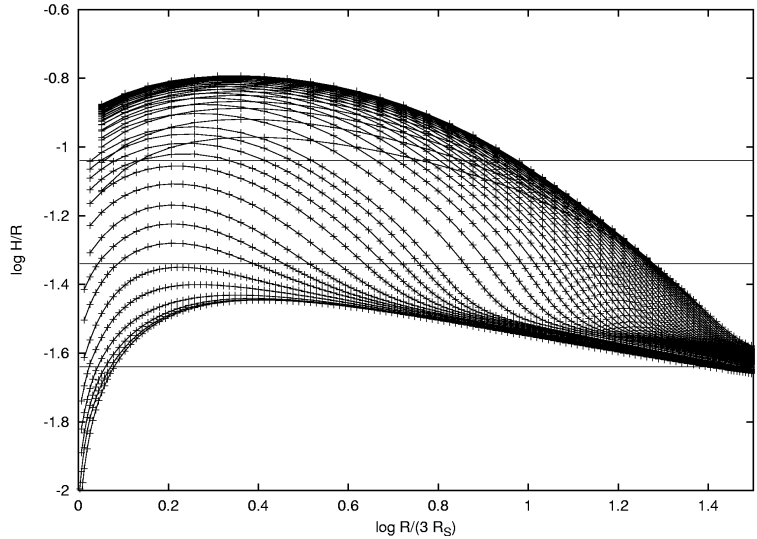}
  \caption{Example for the refinement/coarsening procedure to account
    for the changing scale height in the disc. (Fig. 1 of \citet{2006MNRAS.368..379M}).}
  \label{fig:amr}
\end{figure}

Furthermore we introduced a time-dependent energy equation. This
equation is complemented by the condition of hydrostatic equilibrium
to account for the vertical structure. The poloidal magnetic field
creates a magnetic torque which influences the viscous evolution of
the disc. For the magnetic torque we take the average of the magnetic
field in neighbouring cells to account for the reconnection process,
i.e. we only get a large magnetic torque, if the magnetic field is
sufficiently aligned over some distance.

The disc is a optically thick but geometrically thin disc (see
\citet{1973A&A....24..337S}). Thus the model in the present extension
is only applicable to the so-called high-soft state in X-Ray binaries,
where most of the flux comes from an black-body component. This
component is thought to be produced by the optically thick disc.

There are however still a few problems in this standard accretion disc
picture. All models of accretion discs using the standard model of an
optically thick disc produce limit cycles (i.e. periodic variations in
the luminosity) which occur on much shorter timescales than actually
observed. The only source with limit-cycle behaviour on timescales
comparable to theoretical models is GRS1905+105. All other sources
vary on much longer timescales. 

For our analysis we thus only take the high-luminosity part of the
light-curve as a model for the high/soft state, while we attribute the
low-luminosity part to the yet physically unresolved low/hard state. 

\begin{figure}
  \centering
  \includegraphics[angle=-90,width=0.7\textwidth]{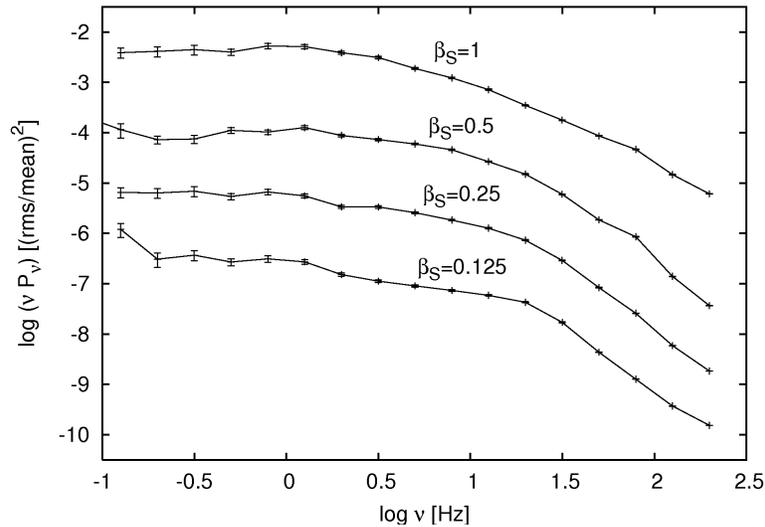}
  \caption{Power density spectra for a 10 M$_\odot$ black hole
    accreting at 0.5 $\dot M_\textrm{Edd}$ with $\alpha=0.1$ for
    different values of $\beta_S$, the strength of the poloidal
    magnetic field compared to the intrinsic magnetic field of the
    disc (Fig. 13 of \citet{2006MNRAS.368..379M}).}
  \label{fig:pds}
\end{figure}

We show sample power spectra from our model in
Fig.~\ref{fig:pds}. Since the observed variability in this state is
fairly small (the integrated rms variability is less
than or around one per cent, see \citet{2006MNRAS.368..379M} for a review of
observational results), we are able to constrain the value of the
maximum allowed energy density of the poloidal magnetic field to be
less than a few per cent of the intrinsic magnetic field density of
the disc, i.e. $\beta_S\leq 0.25$ (Note that the fraction of the
magnetic field energy density is $\beta_S^2$). 

The overall shape of the model power density spectra in
Fig.~\ref{fig:pds} agrees very well with the general shape of observed
power spectra (even in the low/hard state, although then the amplitude
is much higher). Higher-resolution power density spectra however show
more features than just the flat part at low frequencies and then the
decline to higher frequencies. \citet{2003A&A...407.1039P} show that
the power density spectrum of Cyg X-1 can be well modelled by a number
of broad Lorentzian components. This ''fine-structure'' certainly
contains more information about the internal structure of the
accretion flow. 

\section{Conclusions}

We present a extension to the model for the flickering of black hole
accretion disc of \citet{2004MNRAS.348..111K}. Since we take proper
account of the vertical structure and trace the thermal evolution of
the disc, we are able to constrain the energy density of the poloidal
magnetic field to be less than a few per cent of the intrinsic
magnetic field density of the disc. This result, however, strongly depends on
the geometric structure of the accretion flow in this state. 

Our model now can be extended to model the low/hard state variability
as well. Then the X-Ray spectrum shows a powerlaw behaviour. This
powerlaw is thought to be produced in an optically thin corona above
an beneath the accretion disc. To what extent or if at all the
optically thick accretion disc then is truncated at some radius, is
still rather uncertain and yet has to be explored.

  

\bibliographystyle{aipproc}   


\begin{thebibliography}{10}
\expandafter\ifx\csname natexlab\endcsname\relax\def\natexlab#1{#1}\fi
\providecommand{\enquote}[1]{``#1''}
\expandafter\ifx\csname url\endcsname\relax
  \def\url#1{\texttt{#1}}\fi
\expandafter\ifx\csname urlprefix\endcsname\relax\def\urlprefix{URL }\fi
\providecommand{\eprint}[2][]{\url{#2}}

\bibitem[{King} et~al.(2004)]{2004MNRAS.348..111K}
A.~R. {King}, J.~E. {Pringle}, R.~G. {West}, and M.~{Livio}, \emph{\mnras}
  \textbf{348}, 111--122 (2004).

\bibitem[{Mayer} and {Pringle}(2006)]{2006MNRAS.368..379M}
M.~{Mayer}, and J.~E. {Pringle}, \emph{\mnras} \textbf{368}, 379--396 (2006),
  \eprint{astro-ph/0601663}.

\bibitem[{Remillard} and {McClintock}(2006)]{2006astro.ph..6352R}
R.~A. {Remillard}, and J.~E. {McClintock}, \emph{ArXiv Astrophysics e-prints}
  (2006), \eprint{astro-ph/0606352}.

\bibitem[{Uttley} and {McHardy}(2001)]{2001MNRAS.323L..26U}
P.~{Uttley}, and I.~M. {McHardy}, \emph{\mnras} \textbf{323}, L26--L30 (2001).

\bibitem[{Maccarone} and {Coppi}(2002)]{2002MNRAS.336..817M}
T.~J. {Maccarone}, and P.~S. {Coppi}, \emph{\mnras} \textbf{336}, 817--825
  (2002).

\bibitem[{Uttley} et~al.(2005)]{2005MNRAS.359..345U}
P.~{Uttley}, I.~M. {McHardy}, and S.~{Vaughan}, \emph{\mnras} \textbf{359},
  345--362 (2005), \eprint{astro-ph/0502112}.

\bibitem[{Lyubarskii}(1997)]{1997MNRAS.292..679L}
Y.~E. {Lyubarskii}, \emph{\mnras} \textbf{292}, 679--+ (1997).

\bibitem[{Shakura} and {Sunyaev}(1973)]{1973A&A....24..337S}
N.~I. {Shakura}, and R.~A. {Sunyaev}, \emph{\aap} \textbf{24}, 337--355 (1973).

\bibitem[{Livio} et~al.(2003)]{2003ApJ...593..184L}
M.~{Livio}, J.~E. {Pringle}, and A.~R. {King}, \emph{\apj} \textbf{593},
  184--188 (2003).



\bibitem[{Pottschmidt} et~al.(2003)]{2003A&A...407.1039P}
K.~{Pottschmidt}, J.~{Wilms}, M.~A. {Nowak}, G.~G. {Pooley}, T.~{Gleissner},
  W.~A. {Heindl}, D.~M. {Smith}, R.~{Remillard}, and R.~{Staubert}, \emph{\aap}
  \textbf{407}, 1039--1058 (2003).

\end{thebibliography}

\end{document}